\documentclass[12pt]{iopart}

\usepackage{iopams}  
\usepackage{setstack}
\usepackage{graphicx}
\usepackage{subcaption}
\usepackage{siunitx}
\usepackage{color, soul}

\usepackage{lscape}
\usepackage{longtable}
\usepackage{multirow}
\usepackage{makecell}

\graphicspath{{figures-submit-v4/}}
\newcommand\hfigwidth{3.06in}
\newcommand\sfigwidth{4.5in}

\begin{document}

\title[Grav. redshift test of EEP with RA from near Earth to the distance of the Moon]{Gravitational redshift test of EEP with RadioAstron from near Earth to the distance of the Moon}

\author{N V Nunes$^1,$\footnote[1]{Author to whom any correspondence should be addressed.}, N Bartel$^1$, A Belonenko$^2$, G D Manucharyan$^{2,3}$, S M Popov$^2$, V N Rudenko$^2$, L I Gurvits$^{4,5}$, G Cimò$^4$, G Molera Calvés$^6$, M V Zakhvatkin$^7$, M F Bietenholz$^1$}

\address{$^1$ York University, 4700 Keele St, Toronto, ON, M3J 1P3, Canada}
\address{$^2$ Sternberg Astronomical Institute, Lomonosov Moscow State University, Universitetsky pr. 13, 119234 Moscow, Russia}
\address{$^3$ Bauman Moscow State Technical University, ul. Baumanskaya 2-ya, 105005 Moscow, Russia}
\address{$^4$ Joint Institute for VLBI ERIC, Oude Hoogeveensedijk 4, 7991 PD Dwingeloo, the Netherlands}
\address{$^5$ Aerospace Faculty, Delft University of Technology, Delft, the Netherlands}
\address{$^6$ University of Tasmania, Private Bag 37, Hobart 7001, Tasmania, Australia}
\address{$^7$ Keldysh Institute for Applied Mathematics, Russian Academy of Sciences, Miusskaya sq. 4, 125047 Moscow, Russia}

\ead{nvnunes@yorku.ca}

\begin{abstract}
    The Einstein Equivalence Principle (EEP) is a cornerstone of general relativity and predicts the existence of gravitational redshift. We report on new results of measuring this shift with RadioAstron (RA), a space VLBI spacecraft launched into an evolving high eccentricity orbit around Earth with geocentric distances reaching \SI{353000}{km}. The spacecraft and ground tracking stations at Pushchino, Russia, and Green Bank, USA, were each equipped with a hydrogen maser frequency standard allowing a possible violation of the predicted gravitational redshift, in the form of a violation parameter $\varepsilon$, to be measured. By alternating between RadioAstron's frequency referencing modes during dedicated sessions between 2015 and 2017, the recorded downlink frequencies can essentially be corrected for the non-relativistic Doppler shift. We report on an analysis using the Doppler-tracking frequency measurements made during these sessions and find $\varepsilon = (2.1 \pm 3.3)\times10^{-4}$. We also discuss prospects for measuring $\varepsilon$ with a significantly smaller uncertainty using instead the time-domain recordings of the spacecraft signals and envision how $10^{-7}$ might be possible for a future space VLBI mission.
\end{abstract}


\jl{6}
\submitted{}

\maketitle

\section{Introduction}
\label{sec:introduction}

The symmetries that embody the Einstein Equivalence Principle (EEP) underly metric theories of gravitation like general relativity. However, attempts at a quantum description of gravity seem to inevitably lead to violations of the EEP \cite{Will:2014}. A consequence of such a violation might be a departure from the predicted gravitational redshift:

\begin{equation} \label{eqn:gravitational-redshift}
    y_{\text{grav}} \equiv \frac{\nu_{\text{e}} - \nu_{\text{o}}}{\nu_{\text{e}}} = \left(1+\varepsilon\right)\frac{\Delta U}{c^2}
\end{equation}

\noindent where $\nu$ is the frequency of an electromagnetic signal measured at different points within a gravitational field, $\Delta U$ is the difference in gravitational potential at the points, and $\varepsilon$ the violation parameter in the case where identical atomic frequency standards are used by the emitter ($\text{e}$) and observer ($\text{o}$) \cite{Biriukov:2014}. Measuring $\varepsilon$ is the subject of this paper.

The first high-precision laboratory experiments of this type, reaching a relative accuracy of $1\%$, were done in the 1960s by Pound and Rebka \cite{Pound:1960} and later improved by Pound and Snider \cite{Pound:1964}. In 1976, Gravity Probe A (GP-A) \cite{Vessot:1980} was launched on a non-orbital trajectory, with an apogee altitude of \SI{10000}{km}, allowing the gravitational redshift to be measured with an accuracy of $\sigma_\varepsilon = 1.4 \times 10^{-4}$ \cite{Vessot:1989}. More recently, teams utilizing a pair of Galileo global navigation system satellites, which are in elliptical orbits with an eccentricity of $0.16$, were able to refine the measurement of the violation parameter to $\left(0.19 \pm 2.48\right) \times 10^{-5}$ \cite{Delva:2018} and $\left(4.5 \pm 3.1\right) \times 10^{-5}$ \cite{Herrmann:2018} by taking advantage of an $\sim \SI{8500}{km}$ variation in geocentric distance over the satellites' orbits. Optical lattice clocks have also become sufficiently accurate to allow a similar measurement of $\left(1.4 \pm 9.1\right) \times 10^{-5}$ on Earth \cite{Takamoto:2020}. Proposed future experiments may allow these measurements to be further refined by several orders of magnitude \cite{Savalle:2019,Litvinov:2021b}.

In this paper, we present the latest results from a gravity experiment using RadioAstron (RA), the spacecraft element of the Russian-led international space very long baseline interferometry (VLBI) mission, launched in 2011 into a highly eccentric elliptical orbit with geocentric distances under \SI{7000}{km} and as large as \SI{353000}{km} \cite{Kardashev:2013}, comparable to the Earth-Moon distance. The spacecraft carried a VCH-1010 space-qualified hydrogen maser (SHM) frequency standard \cite{Alexandrov:2012}. The mission's two ground tracking stations in Pushchino, Russia (PU) and Green Bank, WV, USA (GB) are also equipped with hydrogen masers (H-masers). First results were published in 2020 \cite{Nunes:2020} based on measurements of the spacecraft's downlink signals made with the Doppler tracking equipment (Doppler frequency measurements) also used for orbit determination but were limited to $\sigma_\varepsilon = 3\%$ by systematics most likely due to the error in compensating for the non-relativistic Doppler shift. In this follow-up paper, we report on an analysis of Doppler frequency measurements where the non-relativistic Doppler shift could be suppressed resulting in a $100$ times reduction in $\sigma_\varepsilon$. This higher accuracy was achieved through a Doppler compensation scheme (DCS) during dedicated downlink sessions in which a combination of 1-way and 2-way links similar to GP-A could be used \cite{Biriukov:2014,Litvinov:2018}. 

\begin{figure}
    \centering
    \includegraphics[width=5.0in]{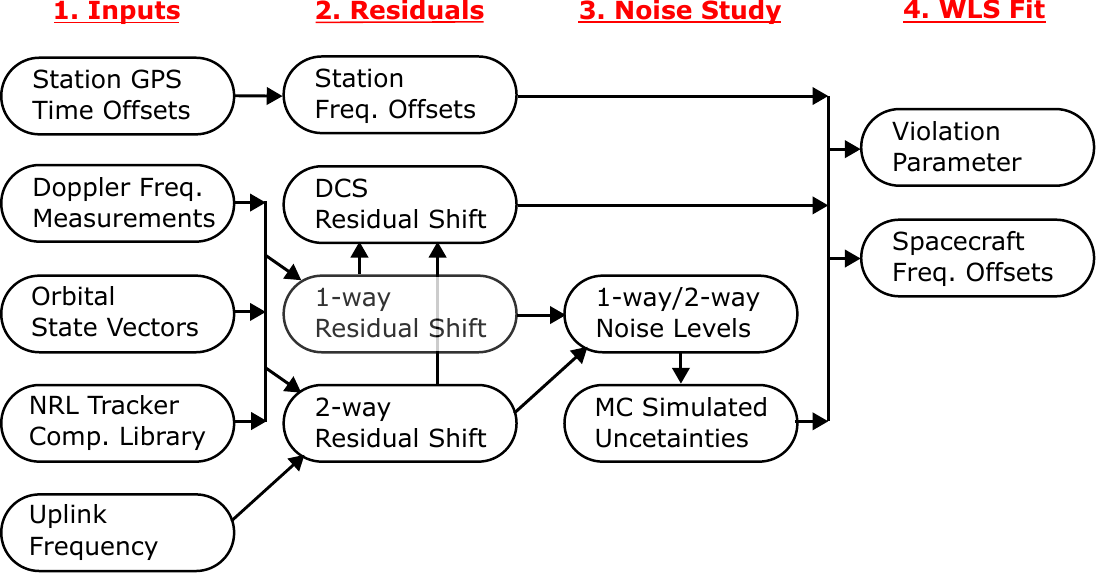}
    \caption{Major steps in the data analysis: 1.~inputs gathered including Doppler frequency measurements, RadioAstron orbital state vectors, and Earth rotation and atmospheric models from the NRL's Tracker Component Library; 2.~residuals computed using the expected frequency shifts; 3.~noise in residuals studied to estimate uncertainties; and 4.~weighted least squares (WLS) fit to estimate the violation parameter, spacecraft frequency offsets and their uncertainties.}
    \label{fig:flow-chart}
\end{figure}

The main steps in the analysis are indicated in \fref{fig:flow-chart}. In the first column are the main inputs: i.~station time offsets relative to GPS time,~ii. Doppler frequency measurements at the ground stations, iii.~orbital state vectors for the spacecraft, iv.~Earth rotation and atmospheric models from the Naval Research Laboratory's (NRL) Tracker Component Library, and v.~log of the instantaneous uplink frequency transmitted to the spacecraft. The next column indicates how the inputs are initially used to compute the frequency offset of the ground station clocks and the residuals when comparing the Doppler frequency measurements to those expected from the state vectors and various models. It is at this stage that the DCS is implemented. The third column indicates that residuals are used to measure noise levels which are used to estimate uncertainties using Monte Carlo techniques. Finally, the last column indicates the use of weighted least squares to simultaneously fit the violation parameter, $\varepsilon$, and the frequency offset of the SHM over time. These steps will be elaborated on in detail in the remainder of this paper which is organized as follows: \sref{sec:doppler-compensation-scheme} describes the DCS and how it was implemented with RA; \sref{sec:measurement-procedures} describes the measurement procedures and the data acquired for the experiment; \sref{sec:data-analysis} describes the data analysis; \sref{sec:estimating-epsilon} describes how $\varepsilon$ and its uncertainty were estimated; \sref{sec:discussion} discusses prospects for further reducing the uncertainty in $\varepsilon$; and \sref{sec:conclusion} provides a brief summary and our conclusions.

\section {Doppler Compensation Scheme}
\label{sec:doppler-compensation-scheme}

\begin{figure}
    \centering
    \includegraphics[width=5.5in]{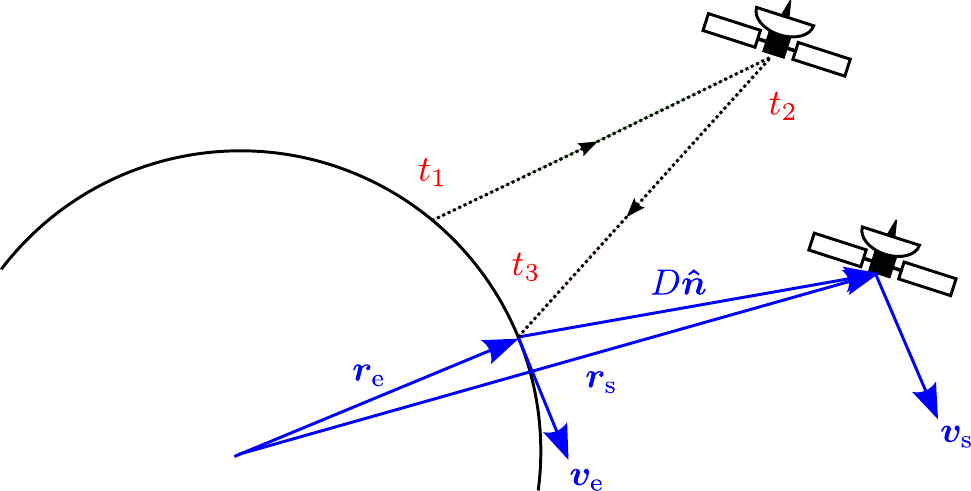}
    \caption{The key times indicated in red are: i.~emission of uplink phase synchronization tone by ground station at $t_1$, ii.~reception of uplink tone by spacecraft and emission of downlink signal at $t_2$ and iii.~reception of downlink signal by ground station at $t_3$. Note that ground station position ($\boldsymbol{r_{\text{e}}}$) and velocity ($\boldsymbol{v_{\text{e}}}$) and spacecraft position ($\boldsymbol{r_{\text{s}}}$) and velocity ($\boldsymbol{v_{\text{s}}}$) are all defined for simplicity at $t_3$, the observation time, as is the range vector ($\boldsymbol{D}\equiv\boldsymbol{r_{s}}-\boldsymbol{r_{e}}=D\boldsymbol{\hat{n}}$).}
    \label{fig:state-vectors}
\end{figure}

RA was equipped with two modes of onboard frequency referencing as shown in \fref{fig:ref-modes}. In the 1-way mode, the observed frequency of the reference tone at \SI{8.4}{GHz} and the carrier signal at \SI{15}{GHz} ($\nu_{\text{1w}}$) experienced a relative frequency shift compared to the nominal or unshifted frequency at the ground tracking station ($\nu_0$):

\begin{figure}
    \centering
    \begin{subfigure}[b]{2.5in}
        \includegraphics[width=2.5in]{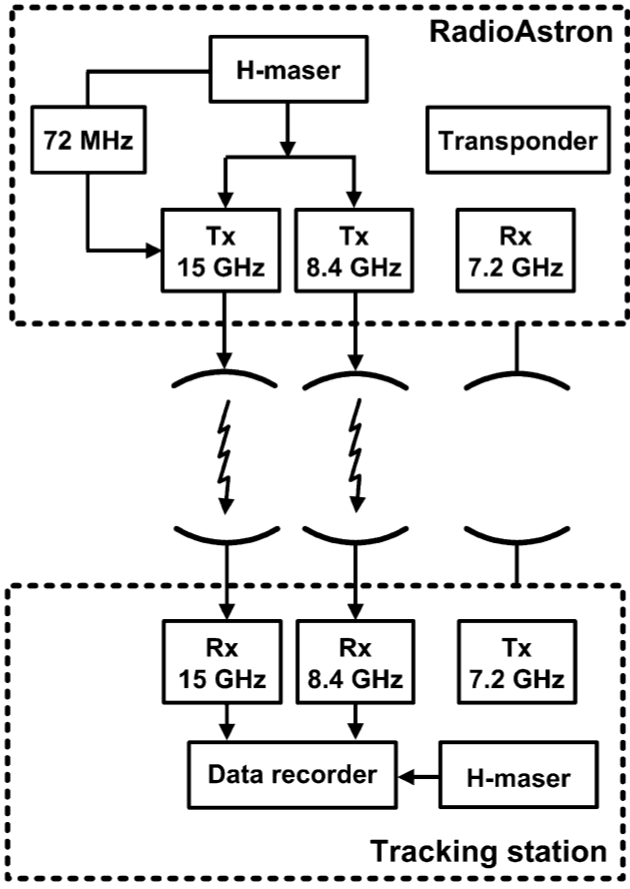}
        \caption{1-way Mode}
        \label{fig:ref-modes-oneway}
    \end{subfigure}
    \begin{subfigure}[b]{2.5in}
        \includegraphics[width=2.5in]{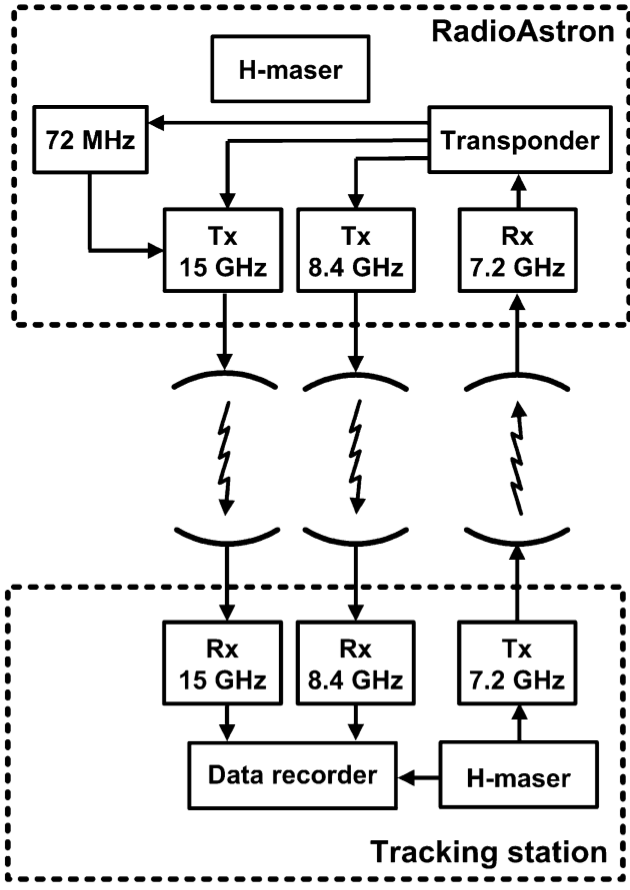}
        \caption{2-way Mode}
        \label{fig:ref-modes-twoway}
    \end{subfigure}
    \caption{RA's frequency referencing modes. (a) In the 1-way mode, the SHM provided the necessary reference to generate the \SI{8.4}{GHz} reference tone and the \SI{15}{GHz} carrier signal. (b) In the 2-way mode, the \SI{7.2}{GHz} reference tone transmitted by the ground tracking station was used in a phase-synchronization loop to provide the onboard frequency reference. Adopted from \cite{Litvinov:2018}.}
    \label{fig:ref-modes}
\end{figure}

\begin{equation} \label{eqn:one-way-frequency-shift-definition}
    y_{\text{1w}} \equiv \frac{\nu_{\text{1w}} - \nu_0}{\nu_0}
\end{equation}

\noindent where $y$ (lowercase) here and hereafter signifies an observed relative frequency shift, $Y$ (uppercase) will be the corresponding expected value and $r$, the residual relative frequency shift or simply `residual shift', will be the difference between them (except in \fref{fig:state-vectors} where $\boldsymbol{r}$ is a position vector). For 1-way, the expected value is given by:

\begin{equation} \label{eqn:one-way-frequency-shift}
    Y_{\text{1w}} = -\frac{\dot{D}}{c} + \frac{\Delta U}{c^2} + \frac{\left|\boldsymbol{v_{\text{e}}}-\boldsymbol{v_{\text{s}}}\right|^{2}}{2c^2} + \frac{\boldsymbol{D}\cdot\boldsymbol{a_{\text{s}}}}{c^2} + Y_{\text{fine,1w}}
\end{equation}

\noindent where $\dot{D}$ is the rate of change of the magnitude of the range vector $\boldsymbol{D}$, also called the range rate, $\boldsymbol{v_{\text{e}}}$ and $\boldsymbol{v_{\text{s}}}$ are respectively the velocities of the ground station and spacecraft, and $\boldsymbol{a_{\text{s}}}$ is the acceleration of the spacecraft. $Y_{\text{fine,1w}}$ includes relativistic Doppler terms at third order or higher in $v/c$ as well as other small effects particularly those due to the troposphere, ionosphere and phase-center motion (PCM) that are considered but omitted here for brevity. Key times and state vectors are as defined in \fref{fig:state-vectors} at $t_3$ (observation time). Vectors at times $t_1$ and $t_2$, when needed, are related to $t_3$ using higher order corrections derived by expansion around $t_3$.

The first order term, the non-relativistic Doppler shift, presents a significant challenge as near perigee it can exceed the gravitational redshift by a factor of $10^4$. Orbital state vectors are typically not accurate enough to estimate $\dot{D}/c$ with sufficient precision and thus dominate the 1-way residual shift: 

\begin{equation} \label{eqn:one-way-residual-frequency-shift}
    r_{\text{1w}} \equiv y_{\text{1w}} - Y_{\text{1w}}
\end{equation}

\noindent For example, $\delta \dot{D} \sim \SI{2}{mm.s^{-1}}$ for RA \cite{Zakhvatkin:2020} which limits the measurement of $y_{\text{grav}}$ to $\sim 1\%$ when using only $r_{\text{1w}}$ \cite{Nunes:2020}. 

In the GP-A experiment, a novel technique was used to suppress the non-relativistic Doppler by taking advantage of a phase-synchronization loop locked to a reference tone uplinked from the ground tracking station to provide a second onboard frequency reference. RA's 2-way mode worked in a similar way. The observed frequency of the downlink signals in this mode ($\nu_{\text{2w}}$) experienced a relative frequency shift compared to the frequency of the uplink tone from the ground tracking station ($\nu_{\text{up}}$) defined as:

\begin{equation} \label{eqn:two-way-frequency-shift-definition}
    y_{\text{2w}} \equiv \frac{\nu_{2w} - F_0 \nu_{\text{up}}}{F_0 \nu_{\text{up}}}
\end{equation}

\noindent where $F_0$ is a multiplier applied by the spacecraft's electronics when generating the downlink frequencies. To second order this shift is given by:

\begin{equation} \label{eqn:two-way-frequency-shift}
    Y_{\text{2w}} = -2\frac{\dot{D}}{c}+2\frac{\left|\boldsymbol{v_{\text{e}}}-\boldsymbol{v_{\text{s}}}\right|^{2}}{c^2}+2\frac{\boldsymbol{D}\cdot\boldsymbol{a_{\text{s}}}}{c^2}-2\frac{\boldsymbol{D}\cdot\boldsymbol{a_{\text{e}}}}{c^2} + Y_{\text{fine,2w}}
\end{equation}

\noindent where $\boldsymbol{a_{\text{e}}}$ is the acceleration of the ground station and $Y_{\text{fine,2w}}$ includes smaller effects, again omitted for brevity as in equation~\eref{eqn:one-way-frequency-shift}. Notice that $y_{\text{2w}}$ does not contain $\frac{\Delta U}{c^2}$. We can again define a residual shift as:

\begin{equation} \label{eqn:two-way-residual-frequency-shift}
    r_{\text{2w}} \equiv y_{\text{2w}} - Y_{\text{2w}}
\end{equation}

\noindent which is also dominated by errors in estimating $\dot{D}/c$. However, with a DCS, both observed relative frequency shifts are combined to form a Doppler compensation scheme relative frequency shift or simply `DCS frequency shift' defined as:

\begin{equation} \label{eqn:dcs-frequency-shift-definition}
    \Delta y \equiv y_{\text{1w}}-\frac{1}{2}y_{\text{2w}}
\end{equation}

\noindent where $\dot{D}/c$ has cancelled and $\frac{\Delta U}{c^2}$ remains as the leading effect in the expected DCS frequency shift:

\begin{eqnarray} \label{eqn:dcs-frequency-shift}
    \Delta Y &\equiv Y_{\text{1w}}-\frac{1}{2}Y_{\text{2w}} \\
    &= \frac{\Delta U}{c^2} - \frac{\left|\boldsymbol{v_{\text{e}}}-\boldsymbol{v_{\text{s}}}\right|^2}{2c^2} + \frac{\boldsymbol{D}\cdot\boldsymbol{a_{\text{e}}}}{c^2} + \Delta Y_{\text{fine}} \nonumber
\end{eqnarray}

\begin{figure}
    \centering
    \begin{subfigure}[b]{\sfigwidth}
        \includegraphics[width=\sfigwidth]{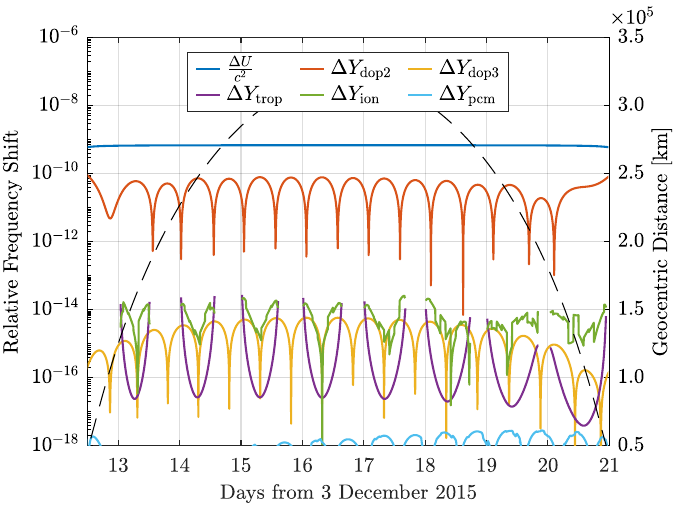}
    \end{subfigure}
    \caption{Relative frequency shifts after implementing the DCS at GB over a particular orbit in December 2015. Dashed line is the spacecraft's geocentric distance and is plotted against the right axes. Note $\Delta Y_{\text{dop2}}$ are the second order relativistic Doppler effects from equation~\eref{eqn:dcs-frequency-shift} and $\Delta Y_{\text{dop3}}$ the third order Doppler terms from equation~\eref{eqn:dcs-frequency-shift-third-order}.}
    \label{fig:model-canc-shifts}
\end{figure}

\noindent The DCS also significantly diminishes the effects of the troposphere, ionosphere and PCM with only differential effects remaining in $\Delta Y_{\text{fine}}$. In order to achieve the desired accuracy, relative frequency shifts as small as $10^{-15}$ must be considered, including terms at third order in $v/c$ which are included in $\Delta Y_{\text{fine}}$:

\begin{eqnarray} \label{eqn:dcs-frequency-shift-third-order}
    \Delta Y_{\text{fine}} &= \Delta Y_{\text{trop}} + \Delta Y_{\text{ion}} + \Delta Y_{\text{pcm}} \\
    &-\frac{D}{c^3}\left[\boldsymbol{D}\cdot\boldsymbol{j_{\text{e}}} + \left(\boldsymbol{v_{\text{e}}}-\boldsymbol{v_{\text{s}}}\right)\cdot\boldsymbol{a_{\text{s}}} + \nabla U_{\text{s}} \cdot \boldsymbol{v_{\text{s}}}\right] \nonumber \\
    &-\frac{\dot{D}}{c^3}\boldsymbol{D}\cdot\boldsymbol{a_{\text{e}}} +\frac{D}{c^3}\left[2\left(\boldsymbol{v_{\text{e}}}-\boldsymbol{v_{\text{s}}}\right)\cdot\boldsymbol{a_{\text{e}}} - \nabla U_{\text{e}} \cdot \boldsymbol{v_{\text{e}}}\right] \nonumber \\
    &-\frac{\dot{D}}{c^3}\left[\Delta U - \frac{1}{2}\left|\boldsymbol{v_{\text{e}}}-\boldsymbol{v_{\text{s}}}\right|^2\right] + O\left(v/c\right)^{4} \nonumber
\end{eqnarray}

\noindent where $\Delta Y_{\text{trop}}$, $\Delta Y_{\text{ion}}$ and $\Delta Y_{\text{pcm}}$ are the differential effects of the troposphere, ionosphere and PCM while $\boldsymbol{j_{\text{e}}}$ is the jerk of the ground station due to Earth's rotation. For RA's orbit, the terms on the last line of equation~\eref{eqn:dcs-frequency-shift-third-order} are negligible as they do not exceed $6\times10^{-16}$. Relativistic Doppler terms are estimated using orbital state vectors for the spacecraft provided by the mission (see~\cite{Zakhvatkin:2020}) along with state vectors for the ground stations computed using the IAU Earth rotation model as implemented in the Tracker Component Library (TCL) by the Naval Research Laboratory (see~\cite{Crouse:2017}). The Earth's gravitational field is modeled using tide-free coefficients from EGM2008 (see~\cite{Pavlis:2012}) with post-glacial rebound, polar motion, solid earth tides, oceanic tides and pole tides added following IERS conventions (see~\cite{IERS2010}). State vectors for the Moon and Sun are computed using JPL's DE421 ephemeris (see~\cite{Folkner:2009}) and used to estimate their tidal effects included in $\Delta U$. $\Delta Y_{\text{trop}}$ is the residual relative frequency shift due to the troposphere estimated using the VMF3 and GRAD models (see~\cite{Landskron:2018vmf3,Landskron:2018grad}). $\Delta Y_{\text{ion}}$ is the residual effect of the ionosphere, mostly due to the difference between uplink and downlink frequencies, estimated using electron densities from CDDIS (see~\cite{Noll:2010}). $\Delta Y_{\text{pcm}}$ includes the PCM effect due to the offset of the spacecraft antenna's phase center from its center of mass (see~\cite{Litvinov:2021}). Also added is the effect due to the significant displacement between the phase center and reference point of the GB ground tracking station. Examples of these relative frequency shifts are plotted over an orbit in \fref{fig:model-canc-shifts}.

Now we introduce our main observable, the Doppler compensation scheme residual relative frequency shift or simply `DCS residual shift':

\begin{eqnarray} \label{eqn:dcs-residual-frequency-shift}
    \Delta r &\equiv \Delta y - \Delta Y \\
             & = r_{\text{1w}} - \frac{1}{2} r_{\text{2w}} \nonumber
\end{eqnarray}

\noindent This observed relative frequency shift only contains unmodelled effects. While H-maser frequency standards may reach or even exceed a relative stability of $10^{-15}$ over thousands of seconds, they are susceptible to a number of systematic effects that cause their frequency to drift over longer times \cite{Vessot:2005}. An offset between each H-maser and Geocentric Coordinate Time (TCG) must be accounted for when comparing observed frequency shifts to prediction. Thus, we define the difference between the SHM and ground tracking station H-maser (GHM) frequency offsets, respectively $h_{\text{s}}$ and $h_{\text{e}}$, as:

\begin{equation} \label{eqn:frequency-bias}
    \Delta h \equiv h_{\text{s}}-h_{\text{e}}
\end{equation}

\noindent This difference will appear in the observed value of $\Delta r$ along with a possible violation of $\frac{\Delta U}{c^2}$ as follows:

\begin{equation} \label{eqn:dcs-residual-frequency-shift-observed}
    \Delta r = \varepsilon\frac{\Delta U}{c^2} + \Delta h
\end{equation}

\noindent The basic approach of this experiment is thus to measure $\Delta r$ using equation~\eref{eqn:dcs-residual-frequency-shift} and fit the resulting observations using a model function based on equation~\eref{eqn:dcs-residual-frequency-shift-observed} to estimate $\varepsilon$ and its uncertainty.

\section{Measurement procedures}
\label{sec:measurement-procedures}

\begin{table}
    \centering
    \caption{Single mode sessions at \SI{8.4}{GHz} used to measure noise levels between March 2015 and December 2017. These $\sim \SI{1}{hr}$ long sessions in which RA performed space VLBI observations employed only one of the two frequency referencing modes from \fref{fig:ref-modes}. Sessions up until mid-2017 were 1-way only while the remainder were 2-way only.}
    \begin{tabular}{| l | c | c | c |}
        \hline
        Type & Station & \makecell{Time Period\\(\footnotesize Days from 1 January 2012)} & Sessions \\
        \hline
        \multirow{2}{*}{\makecell{1-way Mode}} & PU & 1150 - 1948 & $790$ \\
        \cline{2-4}
                                              & GB & 1156 - 1948 & $375$ \\
        \hline
        \multirow{2}{*}{\makecell{2-way Mode}} & PU & 2107 - 2189 &  $45$ \\
        \cline{2-4}
                                              & GB & 2108 - 2189 &  $33$ \\
        \hline
    \end{tabular}
    \label{tab:one-mode-session-counts}
\end{table}

\begin{table}
    \centering
    \caption{Subset of the $199$ interleaved sessions that were used in the experiment. Sessions are $\sim \SI{1}{hr}$ long and have a few or many referencing mode switches. A total of $538$ segments of data could be used to implement the DCS. Sessions with few segments employed a slower switching cycle resulting in segments of longer duration.}
    \begin{tabular}{| l | c | c | c | c | c |}
        \hline
        Type & Station & \makecell{Time Period\\(\footnotesize Days from 1 January 2012)} & Sessions & \makecell{Usable\\Segments} & \makecell{Segment\\Duration} \\
        \hline
        \multirow{2}{*}{\makecell{Few Segments}}  & PU & 1206 - 1943 & $53$ &  $85$ & \SI{213}{s} \\
        \cline{2-6}
                                                  & GB & 1195 - 1935 & $31$ &  $56$ & \SI{216}{s} \\
        \hline
        \multirow{2}{*}{\makecell{Many Segments}} & PU & 1392 - 1939 & $40$ & $368$ & \SI{77}{s} \\
        \cline{2-6}
                                                  & GB & 1426 - 1452 &  $4$ &  $29$ & \SI{77}{s} \\
        \hline
    \end{tabular}
    \label{tab:interleaved-session-counts}
\end{table}

\begin{figure}
    \centering
    \begin{subfigure}[b]{\hfigwidth}
        \includegraphics[width=\hfigwidth]{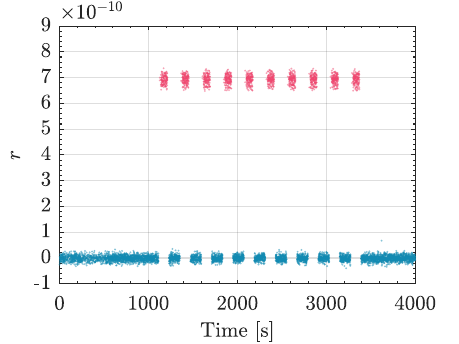}
    \end{subfigure}
    \begin{subfigure}[b]{\hfigwidth}
        \includegraphics[width=\hfigwidth]{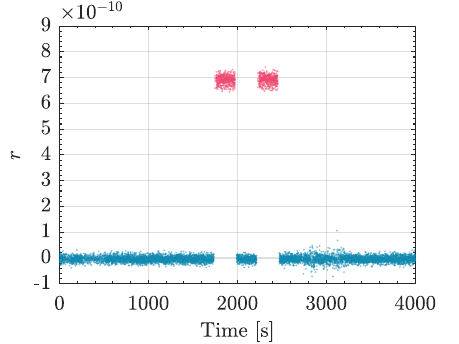}
    \end{subfigure}
    \caption{Residual shifts, $r_{\text{1w}} + \frac{\Delta U}{c^2}$ (red) and $r_{\text{2w}}$ (blue), at \SI{8.4}{GHz} for a session on day 1451 with many mode switches (left) and another session on day 1366 near apogee with only 2 switches (right). Days are counted from 1 January 2012.}
    \label{fig:doppler-residuals}
\end{figure}

Although RA began scientific observations in 2012, measurements for this experiment only started in 2015 and ran until the SHM failed in mid-2017. Most RA sessions involved the real-time downlinking of astronomical space VLBI observations. During these `single mode' sessions, the frequency referencing mode was held fixed and Doppler tracking equipment was used to determined the peak frequency of the spacecraft's signals. This was done for the \SI{8.4}{GHz} reference tone and the \SI{15}{GHz} carrier signal by computing Fourier spectra using \SI{80}{ms} of digitally sampled data with 50\% overlap resulting in a \SI{25}{Hz} measurement rate. A subset of these single mode sessions was used for noise analysis, as will be described later on, and are summarized in \tref{tab:one-mode-session-counts}. In addition, $199$ `interleaved' sessions, each about \SI{1}{hr} long, were dedicated solely to gravitational redshift measurements. During these sessions, RA's frequency referencing mode was switched between 1-way and 2-way modes. Of these sessions, $128$ could be used for this experiment and are summarized in \tref{tab:interleaved-session-counts}. The time series of measurements from an interleaved session can be divided into a series of up to several dozen `segments' over which a particular mode was in use. The interleaving of modes stands out clearly in \fref{fig:doppler-residuals} where the 1-way residual shift with the gravitational redshift added back, $r_{\text{1w}} + \frac{\Delta U}{c^2}$, and the 2-way residual shift, $r_{\text{2w}}$, have been plotted for two sample sessions. Interleaved sessions with many short segments were recorded at a variety of distances while sessions with fewer but longer segments mostly took place close to apogee where the gravitational redshift between the GHM and SHM is at its maximum value. The former provide sensitivity to a possible violation of the gravitational redshift while the latter largely provide sensitivity to the evolution of the frequency offset between the H-masers introduced in the next section.

\section{Data analysis}
\label{sec:data-analysis}

\subsection{Measuring the DCS Residual Shift}
\label{sec:implementing-dcs}

\begin{figure}
    \centering
    \begin{subfigure}[b]{\hfigwidth}
        \includegraphics[width=\hfigwidth]{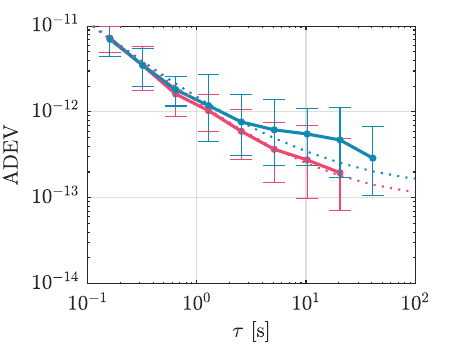}
    \end{subfigure}
    \begin{subfigure}[b]{\hfigwidth}
        \includegraphics[width=\hfigwidth]{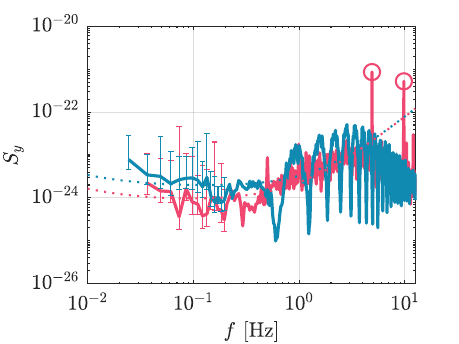}
    \end{subfigure}
    \caption{Example of ADEV (left) and power spectra ($S_y$, right) at \SI{8.4}{GHz} computed using residual shifts $r_{\text{1w}}$ (red) and $r_{\text{2w}}$ (blue) from a typical session. Error bars are 68\% confidence intervals and are omitted at higher frequencies for clarity. Dotted lines correspond to the noise model, fit to the mean $S_y$ from many sessions. Circles indicate interference spikes.}
    \label{fig:unfiltered-residuals-raks13al}
\end{figure}

Values of the DCS residual shift, $\Delta r$, were measured by fitting polynomials to the 1-way and 2-way residual shifts, $r_{\text{1w}}$ and $r_{\text{2w}}$, to interpolate simultaneous relative frequency shifts and apply the DCS. This was done using linear least squares (LLS). In the presence of only white noise, LLS would correctly estimate the interpolation error and correspondingly the uncertainty of $\Delta r$. However, in the presence of non-white noise, such estimates can be significantly biased. To determine the nature of the noise present, the Allan deviation (ADEV as a function of averaging time $\tau$) \cite{Barnes:1971} and power spectral density ($S_y$ as a function of frequency $f$) were computed using $r_{\text{1w}}$ and $r_{\text{2w}}$ and are shown for a typical session in \fref{fig:unfiltered-residuals-raks13al}. Colored noise is evident in the spectrum and can be characterized by spectral index, $\alpha$, where $S_y \propto f^\alpha$. At high frequencies, phase noises such as white phase modulation noise (WPM), with $\alpha = 2$, and flicker phase modulation noise (FPM), with $\alpha = 1$, dominate, while at low frequencies, flicker frequency modulation (FFM) noise, with $\alpha = -1$, begins to dominate above a floor of white frequency modulation (WFM) noise where $S_y$ is constant. In the presence of these noises, LLS error estimates will be biased by the phase noises while the non-stationary nature of FFM will introduce a minimum error (the `flicker noise floor') which cannot be overcome by fitting more data. In the presence of WFM alone, a LLS fit using all the data in a session would result in a single measurement of the DCS residual shift with the smallest possible error. However, due to the flicker noise floor, it is instead advantageous to partition the data and separately fit the frequencies in each part. The error introduced by FFM will be random from fit to fit and will thus allow the effect of the flicker noise floor to be diminished by having multiple independent values of $\Delta r$ per session. In practice, we chose to measure $\Delta r$ at a point in each 1-way segment where the error is expected to be smallest and for which a portion of neighboring 2-way segments could be fit to obtain a simultaneous 2-way mode frequency. A total of $538$ measurements of $\Delta r$ could be made using these segments at \SI{8.4}{GHz} (see \tref{tab:interleaved-session-counts}) with only $519$ of these also being usable at \SI{15}{GHz}.

\subsection{Measuring ground tracking station offsets}
\label{sec:station-frequency-offset}

\begin{figure}
    \centering
    \begin{subfigure}[b]{\hfigwidth}
        \includegraphics[width=\hfigwidth]{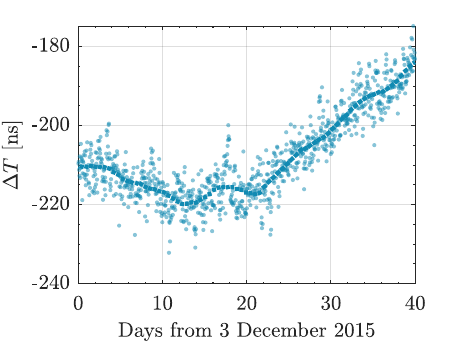}
        \caption{GB GPS Time Offsets}
        \label{fig:gb-gps-offsets}
    \end{subfigure}
    \begin{subfigure}[b]{\hfigwidth}
        \includegraphics[width=\hfigwidth]{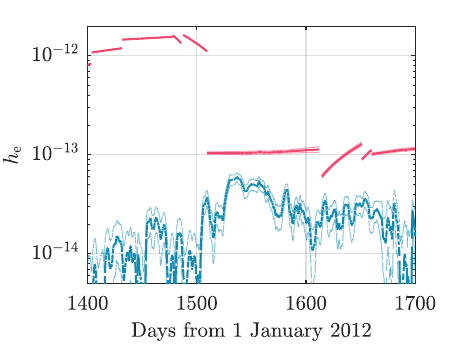}
        \caption{Station Frequency Offsets}
        \label{fig:station-freqency-offsets}
    \end{subfigure}
    \caption{(a) Time offset measurements ($\Delta T$) between GB local time and GPS time with the superimposed trend, given by non-overlapping regions from numerous sub-interval fits, showing both the effect of a systematic H-maser drift and random run noise. (b) Sample range of station frequency offsets, $h_{\text{e}}$, for PU (red) and GB (blue) with confidence intervals (ligher lines) only visible when $\sigma_{h_{\text{e}}}$ is sufficienty large relative to $h_{\text{e}}$.}
    \label{fig:station-frequency-offsets-both}
\end{figure}

PU and GB ground tracking stations are equipped with GPS receivers that allow their local clocks to be steered such that they remain within a maximum time offset relative to GPS time. As GPS time is itself steered to follow Terrestrial Time (TT), a time series of offsets between the local clock and GPS time ($\Delta T$) allow the relative frequency offset of the local clock relative to TT, $h_{\text{e}}$ or station frequency offset, to be estimated using:

\begin{equation} \label{eqn:frequency-offset-station}
    h_{\text{e}}\left(t\right) = \frac{\text{d}}{\text{dt}}\Delta T\left(t\right)-\frac{\mu_\oplus}{c^2} \frac{H_{\text{e}}}{r_{\text{g,e}}^2}
\end{equation}

\noindent where the derivative is taken with respect to TT, $\mu_\oplus$ is the standard gravitational parameter of Earth while $H_{\text{e}}$ and $r_{\text{g,e}}$ are the orthometric height and geoid radius at the station's location. The second term accounts for the station not being on the geoid where TT is defined. Prior to fitting, the time series is resampled into uniform hourly measurements. A portion of the resampled time series for GB is shown in \fref{fig:gb-gps-offsets}. Note the two dominant types of noise present: (1) white noise from the GPS receiver and (2) random run phase noise due to the random walk frequency noise of the GHM. These are superimposed on the systematic drift of the maser's offset which, to first-order, is linear in frequency and therefore quadratic in the time offset. However, when an H-maser is disturbed, sudden changes in drift are possible, as seen near day 20 in \fref{fig:gb-gps-offsets}. The first step in estimating the station frequency offset, $h_{\text{e}}$, is to divide the time series into intervals over which the H-maser was undisturbed. This was done using operator logs from GB and manually by inspecting the PU time-series looking for discontinuities. Within each interval, a single quadratic fit of the time series would be appropriate if only white noise were present. However, due to the random run, each interval must be divided into sub-intervals which are as long as possible to minimize the effect of white noise, but over which the random run does not dominate. The optimal length of these sub-intervals was determined by measuring the white noise level and using GHM specifications for the frequency random walk noise level. Overlapping quadratic fits of the optimal length were done and the non-overlapping regions with the lowest uncertainty from each fit were used to produce a time-series of frequency offsets, a range of which are plotted in \fref{fig:station-freqency-offsets}. Finally, the uncertainties of the frequency offsets measured using this approach were determined using Monte Carlo simulation with randomly generated noise (see \cite{Zucca:2005} for a general approach to generating clock noise) according to the determined noise levels. These uncertainties, $\sigma_{h_{\text{e}}}$, are also plotted in \fref{fig:station-freqency-offsets}.

\subsection{Measuring spacecraft frequency offset and $\varepsilon$}
\label{sec:frequency-offsets}

\noindent As the station frequency offsets, $h_{\text{e}}$, can be measured independently using the method described in \sref{sec:station-frequency-offset}, we consider it an observable along with the DCS residual shift, $\Delta r$, and so modify equation~\eref{eqn:dcs-residual-frequency-shift-observed} to the following: 

\begin{equation} \label{eqn:dcs-residual-frequency-shift-final}
    \Delta r + h_{\text{e}} = \varepsilon\frac{\Delta U}{c^2} + h_{\text{s}}
\end{equation}

\begin{figure}
    \centering
    \begin{subfigure}[b]{\sfigwidth}
        \includegraphics[width=\sfigwidth]{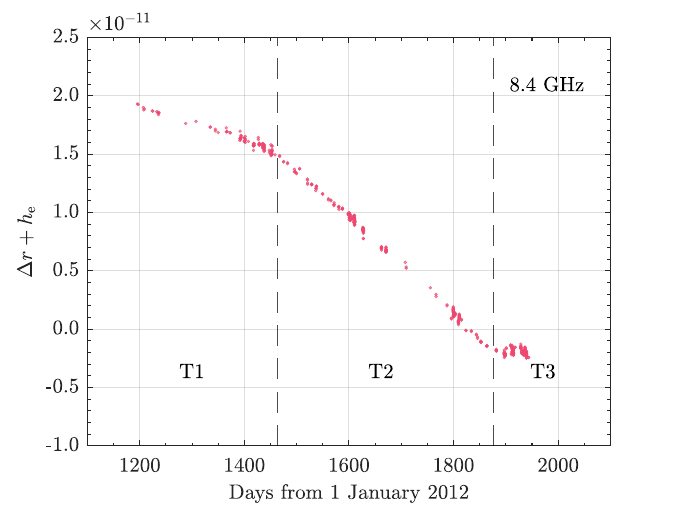}
    \end{subfigure}
    \begin{subfigure}[b]{\sfigwidth}
        \includegraphics[width=\sfigwidth]{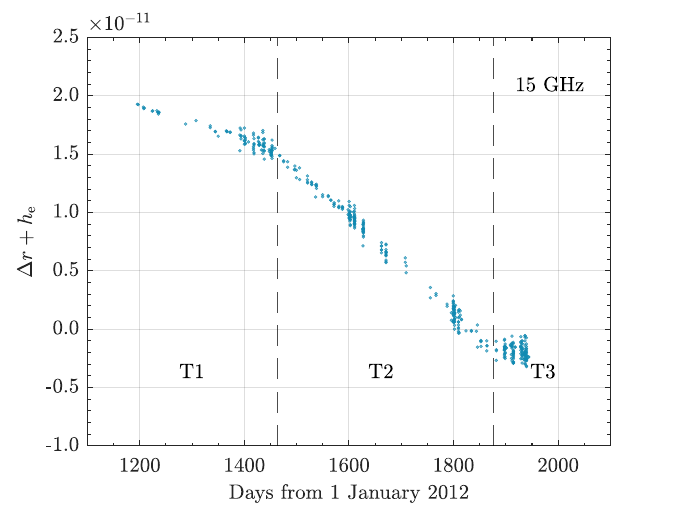}
    \end{subfigure}
    \caption{The main observables in the experiment, $\Delta r + h_{\text{e}}$ from equation~\eref{eqn:dcs-residual-frequency-shift-final}, to which the model function, $E\left(\frac{\Delta U}{c^2}, t, \boldsymbol{\beta}\right)$, was fit to obtain $\varepsilon$ and the spsacecraft frequency offset, $h_{\text{s}}$. Each of the $531$ points at \SI{8.4}{GHz} and $473$ points at \SI{15}{GHz} correspond to a 1-way segment for which the DCS could be implemented at either GB or PU. The trend is dominated by the linear drift of $h_{\text{s}}$ over 3 intervals (T1, T2, T3) with boundaries near days 1464 and 1875 (dashed lines). The larger scatter at \SI{15}{GHz} starting near day 1400 is due to PU points which have more noise. Day $1200$ corresponds to 15 April 2015.}
    \label{fig:model-residual-frequency-shift}
\end{figure}

\noindent Values of $\Delta r + h_{\text{e}}$ plotted versus time appear in \fref{fig:model-residual-frequency-shift}. The nearly linear drift of the spacecraft relative frequency offset, $h_{\text{s}}$, is apparent. Discriminating $h_{\text{s}}$ from the effect of $\varepsilon$ requires long intervals over which $h_{\text{s}}$ evolves linearly and over which the range of gravitational redshifts is as large as possible. While the data cover an impressive range of redshifts, $\Delta y_{\text{grav}} = 1.6 \times 10^{-10}$ corresponding to a distance range of \SI{320000}{km} from \SI{26000}{km} to \SI{346000}{km}, a full $2/3$ of points are near apogee within $20\%$ of the maximum value of the gravitational redshift. This has the effect of strongly correlating $\varepsilon$ with the initial value of the spacecraft frequency offset, $h_{\text{s}}$, and requires that both be fit simultaneously.

In \fref{fig:model-residual-frequency-shift}, we see the slope of $h_{\text{s}}$ changing suddenly on two occasions. We thus divide the data into three intervals (T1, T2, T3 arranged chronologically) with boundaries at days $1464.0$ and $1875.5$ where days are counted starting on 1 January 2012. Sensitivity to this choice is discussed in \sref{sec:systematic-error}. Within each interval, we assume a linear drift and so define the following model function:

\begin{equation} \label{eqn:residual-frequency-shift-model}
    E\left(\frac{\Delta U}{c^2}, t, \boldsymbol{\beta}\right) = \varepsilon \frac{\Delta U}{c^2} + \left[\sum_{i=1}^3 \Pi\left(t,i\right) \left(a_i + b_i\:t\right)\right] 
\end{equation}

\noindent where the factor $\Pi\left(t,i\right)$ is $1$ if $t$ lies within interval $Ti$ and $0$ otherwise. The parameter vector, $\boldsymbol{\beta}$, includes $\varepsilon$ and the `$h_{\text{s}}$ parameters', $a_i$ and $b_i$, that account for the drift of the SHM frequency offset. Having three $a_i$ parameters instead of just one overall constant, allows for discontinuities at the boundary between intervals, which reduces the sensitivity of $E\left(\frac{\Delta U}{c^2}, t, \boldsymbol{\beta}\right)$ to the precise choice of boundary times. In total, this model function has $7$ parameters including $\varepsilon$ for fitting to the measurements of $\Delta r + h_{\text{e}}$.

\section{Estimating the violation parameter, $\varepsilon$}
\label{sec:estimating-epsilon}

\subsection{Unweighted fit}
\label{sec:unweighted-fit}

\begin{figure}
    \centering
    \begin{subfigure}[b]{\hfigwidth}
        \includegraphics[width=\hfigwidth]{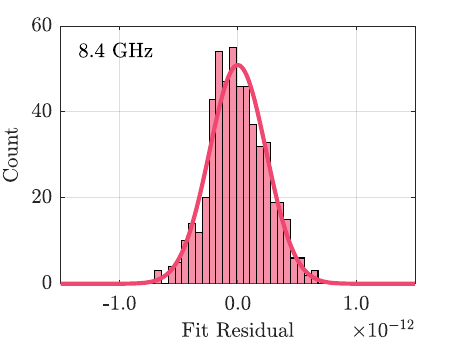}
    \end{subfigure}
    \begin{subfigure}[b]{\hfigwidth}
        \includegraphics[width=\hfigwidth]{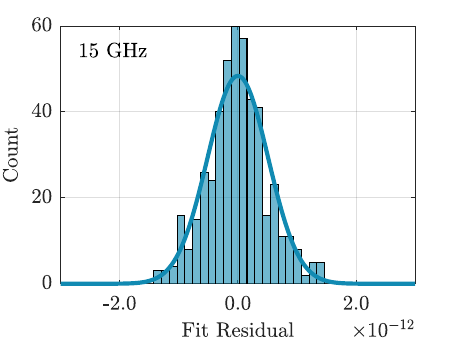}
    \end{subfigure}
    \begin{subfigure}[b]{\hfigwidth}
        \includegraphics[width=\hfigwidth]{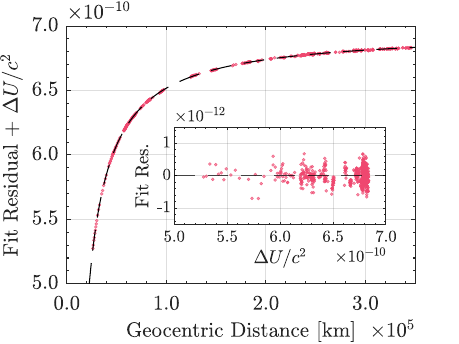}
    \end{subfigure}
    \begin{subfigure}[b]{\hfigwidth}
        \includegraphics[width=\hfigwidth]{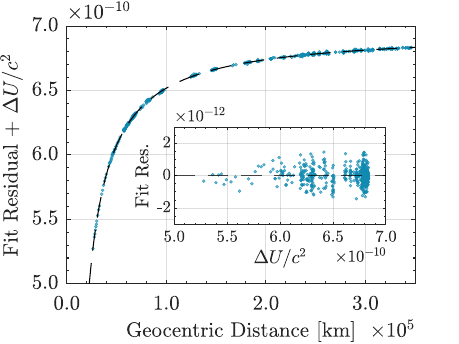}
    \end{subfigure}
    \caption{Top are distributions of residuals from fitting the model function, $E\left(\frac{\Delta U}{c^2}, t, \boldsymbol{\beta}\right)$, to $\Delta r + h_{\text{e}}$ with overlayed normal distributions (solid line) centered at zero with standard deviations corresponding to the RMS of the residuals. Bottom are the same fit residuals but with gravitational redshift added and plotted versus distance. Insets are the fit residuals versus gravitational redshift.}
    \label{fig:method3c-residuals}
\end{figure}

First, the model function, $E\left(\frac{\Delta U}{c^2}, t, \boldsymbol{\beta}\right)$, was fit to the data in \fref{fig:model-residual-frequency-shift} using unweighted LLS. Following a preliminary fit, $7$ points ($1\%$) at \SI{8.4}{GHz} and $46$ points ($9\%$) at \SI{15}{GHz} were excluded using a $3$ scaled median-absolute-deviations from the median criterion leaving $N_{\text{8.4}} = 531$ and $N_{\text{15}} = 473$ points. Fit residuals are shown in \fref{fig:method3c-residuals} and have root-mean-square (RMS) values of $\text{RMS}_{\text{8.4}}=2.4\times10^{-13}$ and $\text{RMS}_{\text{15}}=5.1\times10^{-13}$. The values of $\varepsilon$ from the unweighted fit are: $\varepsilon_{\text{unw,8.4}} = \left(2.2 \pm 3.4_{\text{stat}}\right) \times 10^{-4}$ and $\varepsilon_{\text{unw,15}} = \left(1.5 \pm 7.6_{\text{stat}}\right) \times 10^{-4}$. However, given that segments are not all of the same size and noise levels varied throughout the experiment, a weighted fit incorporating the expected uncertainty in each measurement of $\Delta r + h_{\text{e}}$ is more appropriate.

\subsection{Weighted fit}
\label{sec:weighted-fit}

The uncertainty of the measurements, $\sigma$, was used as a weighting factor and, based on equations~\eref{eqn:dcs-residual-frequency-shift} and \eref{eqn:dcs-residual-frequency-shift-final}, includes the statistical uncertainties $\sigma_{r_{\text{1w}}}$, $\sigma_{r_{\text{2w}}}$ and $\sigma_{h_{\text{e}}}$ from the fits described in sections~\ref{sec:implementing-dcs} and \ref{sec:station-frequency-offset}. The uncertainty in the expected DCS frequency shift, $\Delta Y$, was ignored as it is either too small or not statistical in nature. This will be further discussed along with estimating the systematic error in \sref{sec:systematic-error}. Combining the uncertainties in quadrature gives the total uncertainty:

\begin{equation} \label{eqn:residual-frequency-shift-uncertainty}
    \sigma^2 = \sigma_{r_{\text{1w}}}^2 + \frac{1}{4}\sigma_{r_{\text{2w}}}^2 + \sigma_{h_{\text{e}}}^2
\end{equation}

\begin{figure}
    \centering
    \begin{subfigure}[b]{\hfigwidth}
        \includegraphics[width=\hfigwidth]{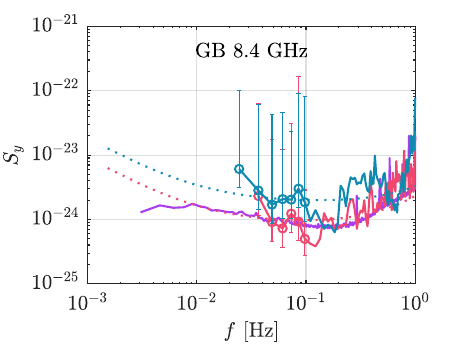}
    \end{subfigure}
    \begin{subfigure}[b]{\hfigwidth}
        \includegraphics[width=\hfigwidth]{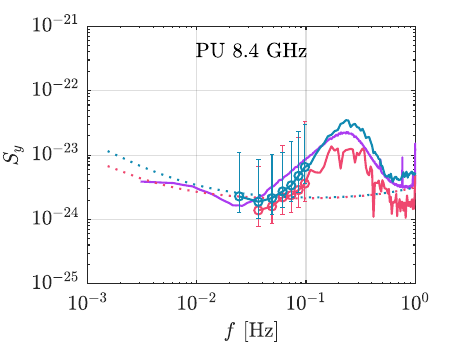}
    \end{subfigure}
    \caption{Mean power spectra of the residual shifts $r_{\text{1w}}$ (red) and $r_{\text{2w}}$ (blue) at \SI{8.4}{GHz} computed using segments from interleaved sessions. Left is the mean of GB sessions and right is that of PU sessions between 7 May 2015 and 3 June 2016. Dotted lines are corresponding noise models. The mean spectra from 1-way only sessions are plotted in magenta. Some error bars are omitted for legibility. Note, the presence of additional noise above \SI{0.02}{Hz} in PU spectra which is discussed in the text.
    }
    \label{fig:noisefit-spectrum}
\end{figure}

\noindent Due to the presence of non-white noise, the statistical uncertainties could not be estimated from their corresponding LLS fits. Instead, they were estimated as confidence intervals using Monte Carlo simulations of colored noise generated using models matched to observed noise power (see $S_y$ in \sref{sec:implementing-dcs}). These models include four power-law noise components (WPM, FPM, WFM and FFM) and were fit to the mean $S_y$ from many sessions (see \fref{fig:noisefit-spectrum}). Only single mode sessions, all 1-way or 2-way mode, were used to compute mean $S_y$ as their longer stretches of data are better suited to measuring non-stationary noises. The presence of additional noise at intermediate frequencies above \SI{0.02}{Hz} in PU spectra after 7 May 2015 obscures the noise floor relevant at longer averaging times. Therefore, we used the FFM noise level fit to sessions between 24 February 2015 and 7 May 2015 and adjusted the WFM noise level to match that observed in the sessions coming after. As 2-way only sessions were not performed before additional noise appears in PU spectra, the 2-way noise level for PU could not be fit. Instead, FFM noise power was assumed to be twice that in 1-way, which is approximately the same ratio seen between 2-way and 1-way power at GB. Monte Carlo simulation showed that FFM and, to a lesser extent, WFM dominate the error. Care was taken to generate FFM noise with the appropriate characteristics using an $\textrm{ARFIMA}\left(1,0.5,0\right)$ stochastic model following the approach of \cite{Xu:2019}. Using 1000 simulations per session, the uncertainties in the 1-way and 2-way residual shifts, $\sigma_{r_{\text{1w}}}$ and $\sigma_{r_{\text{2w}}}$, were found to respectively contribute $74\%$ and $26\%$, on average, to the total uncertainty while the contribution of the uncertainty in the station frequency offset, $\sigma_{h_{\text{e}}}$, is negligible. The mean total uncertainties across all segments are $\overline{\sigma}_{\text{8.4}} = 2.1 \times 10^{-13}$ and $\overline{\sigma}_{\text{15}} = 2.5 \times 10^{-13}$. 

Using the total uncertainties as weighting factors when fitting equation~\eref{eqn:residual-frequency-shift-model}, the following values of $\varepsilon$ were obtained: $\varepsilon_{\text{8.4}} = \left(2.1 \pm 3.3_{\text{stat}}\right) \times 10^{-4}$ and $\varepsilon_{\text{15}} = \left(0.7 \pm 7.6_{\text{stat}}\right) \times 10^{-4}$. These results are very similar to those from the unweighted fits. The weighted fits have chi-square per degree of freedom of $\chi^2_{\nu\text{,8.4}}=1.1$ and $\chi^2_{\nu\text{,15}}=4.4$. The former suggests that the weighting factors determined using Monte Carlo methods account for nearly all the scatter in $\Delta r + h_{\text{e}}$ at \SI{8.4}{GHz}. In contrast, the larger $\chi^2_\nu$ at \SI{15}{GHz}, which is evident from $\text{RMS}_{\text{15}}$ being significantly larger than $\overline{\sigma}_{15}$, implies that the estimated uncertainty at \SI{15}{GHz} is too low. This is not surprising given that we could not directly fit noise levels in this band and instead resorted to using the lower noise levels from \SI{8.4}{GHz}. The statistical uncertainties of $\varepsilon$ have been adjusted so they correspond to a $\chi^2_{\nu}$ of unity.

\subsection{Check on statistical error}
\label{sec:check-statistical-error}

We can compare $\sigma_\varepsilon^{\text{stat}}$ to what is expected given the mean variation in gravitational redshift ($\overline{\Delta y}_{\text{grav}} \sim 2.6\times10^{-11}$), the RMS of the fit residuals, the number of points given in \sref{sec:unweighted-fit} and the mean correlation between $\varepsilon$ and the $h_{\text{s}}$ parameters ($\overline{\rho}_{8.4} = -21\%$) as follows:

\begin{equation}
    \hat{\sigma}_{\varepsilon\text{,8.4}} \sim \frac{1}{\overline{\Delta y}_{\text{grav}}} \sqrt{\frac{\text{RMS}_{8.4}^2}{N_{8.4}}\left(1+2\overline{\rho}_{8.4}\right)} = 3.1 \times 10^{-4}
\end{equation}

\noindent Using this value, we find $\sigma_{\varepsilon}^{\text{stat}} / \hat{\sigma}_\varepsilon = 1.07$ at \SI{8.4}{GHz} which is fairly close to unity. For \SI{15}{GHz}, this check is not useful since noise levels could not be directly measured. Thus we conclude, at least at \SI{8.4}{GHz}, that the statistical uncertainty is a reliable estimate.

\subsection{Systematic error}
\label{sec:systematic-error}

We tested our analysis technique for a bias when measuring $\varepsilon$, particularly towards $\varepsilon = 0$. By assuming a non-zero violation in the presence of simulated noise we confirmed that our estimate of $\varepsilon$ is unbiased. Further, as mentioned in \sref{sec:frequency-offsets}, these tests confirmed that $h_{\text{s}}$ cannot be measured independently from $\varepsilon$ using the same data set without a possible violation being suppressed. This shows that our overall approach of fitting $\varepsilon$ and $h_{\text{s}}$ simultaneously is necessary.

To determine the contributions to systematic uncertainty, we considered three effects. First, we studied the effect of the interference spikes that appear in 1-way mode spectra (see \fref{fig:unfiltered-residuals-raks13al}). A shift, $\Delta \varepsilon_{\text{filt}}$, resulted when passing the DCS residual shift through a \SI{3}{Hz} Butterworth lowpass filter of order $8$ to remove the spikes prior to fitting. As we could not ascertain which result is more likely to be correct, we conservatively include these shifts, $\Delta \varepsilon_{\text{filt,8.4}}=3\times10^{-5}$ and $\Delta \varepsilon_{\text{filt,15}}=4.2\times10^{-4}$, in the error.

Second, the uncertainty of the specific boundaries between intervals T1, T2, and T3 (see \fref{fig:model-residual-frequency-shift}) were studied. Alternate boundaries between T1 and T2 as well as between T2 and T3 were tried corresponding to where the $1\sigma$ confidence intervals of the fits on either side of the boundary meet, namely at days $1470$ and $1890$ respectively. The differences in the fit values, $\Delta \varepsilon_{\text{bound,8.4}}=2\times10^{-5}$ and $\Delta \varepsilon_{\text{bound,15}}=5\times10^{-5}$, are also added to the error. 

Third, the terms in equation~\eref{eqn:dcs-frequency-shift} larger than the uncertainty in the DCS frequency shift are $\frac{\Delta U}{c^2}$, the second order relativistic Doppler terms and the station frequency offset. As described in \sref{sec:station-frequency-offset}, the uncertainty in the station frequency offset was estimated and is included in the weights and, therefore, the statistical error. For the other terms, the main sources of error are the uncertainty in the spacecraft state vectors ($\delta r \sim \SI{200}{m}$ and $\delta v \sim \SI{2}{mm.s^{-1}}$ \cite{Zakhvatkin:2020}), and the position of PU's reference point ($\delta r < \SI{10}{m}$). The error introduced in the expected DCS frequency shift by these uncertainties does not exceed $1 \times 10^{-15}$, even for the closest perigee session for which the error would be the largest.

Systematic thermal and magnetic effects on the GHMs contribute to the station frequency offset and thus are taken into account. Ground testing of the SHM showed a thermal sensitivity of $\Delta f/f = \pm 5\times10^{-15} / ^\circ\text{C}$ and magnetic field sensitivity of $\Delta f/f = \pm 2\times10^{-14} / \text{G}$. During observing sessions, the thermal management system on board RA maintained the SHM temperature with an accuracy of $1 ^\circ\text{C}$. The resulting random frequency shift is therefore expected to be much smaller than the estimated uncertainty in the residual shift, $\overline{\sigma}_{\text{8.4}}$, and can be neglected. Similarly, at the distances of RA's orbit, the Earth's magnetic field is sufficiently weak ($\ll \SI{0.1}{G}$) that effects due to its variation are also negligible.

The systematic error contributions, also listed in \tref{tab:systematic-error}, are added in quadrature to yield total systematic uncertainties of $\sigma^{\text{sys}}_{\varepsilon\text{,8.4}} = 0.5 \times 10^{-4}$ and $\sigma^{\text{sys}}_{\varepsilon\text{,15}} = 4.3 \times 10^{-4}$.

\begin{table}
    \centering
    \caption{Sources of systematic error: (1) variation in $\varepsilon$ due to filtering interference, (2) variation in $\varepsilon$ due to varying SHM frequency offset interval boundaries, and (3) error due to the uncertainty in the expected DCS frequency shift, $\sigma_{\Delta Y}$, estimated by dividing that uncertainty by $\overline{\Delta y}_{\text{grav}} = 2.6\times10^{-11}$.}
    \begin{tabular}{| l | c | c |}
        \hline
        \multirow{2}{*}{\makecell{Source}} & \multicolumn{2}{c|}{Contributions $\times 10^{-4}$} \\
        \cline{2-3}
        & \SI{8.4}{GHz} & \SI{15}{GHz} \\
        \hline
        1. $\Delta \varepsilon_{\text{filt}}$  &  $0.3$ &  $4.2$ \\ 
        2. $\Delta \varepsilon_{\text{bound}}$ &  $0.2$ &  $0.5$ \\
        3. $\sigma_{\Delta Y} < 10^{-15}$ & $<0.4$ & $<0.4$ \\
        \hline
        \hline
        $\sigma_\varepsilon^\text{sys}$ & $0.5$ & $4.3$ \\
        \hline
    \end{tabular}
    \label{tab:systematic-error}
\end{table}

\subsection{Sensitivity study}
\label{sec:sensitivity-study}

In \tref{tab:epsilon-fit-results}, we summarize our results using different subsets of the data with and without weighting. Measuring $\varepsilon$ using only T2, the longest interval, results in a correlation of $72\%$ between $\varepsilon$ and the constant parameter in the frequency offset. By combining all three frequency offset intervals in a single fit, not only is the uncertainty in $\varepsilon$ reduced, but so is the correlation falling by $6\%$ in the case of T2 but by about $24\%$ in the case of the other two intervals. The majority of sessions with numerous switches, performed nearer Earth, were recorded at PU. The addition of GB data makes essentially no difference at \SI{8.4}{GHz}, indicating that the GB data are consistent with PU, but also that the lower noise levels at PU drive the result. The reverse is true at \SI{15}{GHz}, where the additional noise at PU is partially offset by the inclusion of GB data. All the results, including those from unweighted fits, are broadly consistent with each other within the uncertainties. Furthermore, we find in all cases $\varepsilon$ is consistent with zero within $1.1\thinspace\sigma_\varepsilon^{\text{stat}}$.

\begin{table}
    \centering
    \caption{The result of fitting $\varepsilon$ using different portions of the data set with $N$ points both using and not using weights derived from simulation.}
    \begin{tabular}{| l | l | c | c | c | c | c |}
        \hline
        \multirow{2}{*}{\makecell{Station}} & \multirow{2}{*}{\makecell{Interval}} & \multirow{2}{*}{\makecell{Weighted}} & \multicolumn{2}{c|}{\SI{8.4}{GHz}} & \multicolumn{2}{c|}{\SI{15}{GHz}} \\
        \cline{4-7}
        & & & $N$ & $\varepsilon \pm \sigma_\varepsilon^{\text{stat}} \times 10^{-4}$ & $N$ & $\varepsilon \pm \sigma_\varepsilon^{\text{stat}} \times 10^{-4}$ \\
        \hline                                                  
        PU, GB & T1, T2, T3 & Yes & $531$                       
                                  & $\phantom{-}2.1 \pm 3.3$    
                                  & $474$                       
                                  & $\phantom{-}0.7 \pm 7.6$ \\ 
        PU, GB & T2         & Yes & $268$                       
                                  &           $-3.9 \pm 3.6$    
                                  & $232$                       
                                  &           $-1.3 \pm 8.6$ \\ 
        PU     & T1, T2, T3 & Yes & $447$                       
                                  & $\phantom{-}0.9 \pm 3.3$    
                                  & $405$                       
                                  &           $-1.0 \pm 9.7$ \\ 
        \hline
        PU, GB & T1, T2, T3 & No  & $531$                       
                                  & $\phantom{-}2.2 \pm 3.4$    
                                  & $473$                       
                                  & $\phantom{-}1.5 \pm 7.6$ \\ 
        PU, GB & T2         & No  & $268$                       
                                  & $          -2.9 \pm 3.8$    
                                  & $233$                       
                                  & $          -1.5 \pm 8.9$ \\ 
        PU     & T1, T2, T3 & No  & $447$                       
                                  & $\phantom{-}1.6 \pm 3.4$    
                                  & $404$                       
                                  & $\phantom{-}3.0 \pm 9.7$ \\ 
        \hline
    \end{tabular}
    \label{tab:epsilon-fit-results}
\end{table}

\subsection{Final results}
\label{sec:final-results}

Combining our estimates for $\varepsilon$ from the weighted fit and its uncertainty we arrive at the following results:
$\varepsilon_{\text{8.4}} = \left(2.1 \pm 3.3_{\text{stat}} \pm 0.5_{\text{sys}}\right) \times 10^{-4}$ and $\varepsilon_{\text{15}} = \left(0.7 \pm 7.6_{\text{stat}} \pm 4.3_{\text{sys}}\right) \times 10^{-4}$. For a final result, we considered combining the measurements from the two frequency bands. However, the noise in the two bands appears strongly correlated during GB sessions and at least partially correlated during many of the PU sessions. This implies that the measurements at \SI{8.4}{GHz} and \SI{15}{GHz} cannot be considered statistically independent. The result at \SI{8.4}{GHz} is favored since its $\chi^2_\nu$ being close to unity confirms that our weighting scheme derived from colored noise simulations is reliable. Using the \SI{8.4}{GHz} result and combining its statistical and systematic uncertainties in quadrature, we arrive at a final estimate for the violation parameter: 
\[
    \varepsilon = \left(2.1 \pm 3.3\right) \times 10^{-4}
\]

\section{Discussion}
\label{sec:discussion}

Tests of the EEP are considered an important, if not essential, probe of metric theories of gravity \cite{Damour:2012}, with measuring the gravitational redshift being one of the classical tests of general relativity. Our measurements were made with the space VLBI RA mission which was not primarily designed for a gravitational redshift test. In particular, we were limited by the lack of simultaneous downlink signals in the 1-way and 2-way referencing modes and the limited observation time allocated to the experiment. The flicker noise floor of the online frequency measurements by the Doppler tracking equipment is an order of magnitude higher than that of GP-A and almost $35$ times what was determined in the laboratory for the SHM prior to launch. Nevertheless, the mission allowed an accurate measurement of the gravitational redshift from near Earth to almost the distance of the Moon where it asymptotically approaches its maximum relative to Earth's surface (see bottom plots in \fref{fig:method3c-residuals}). 

In addition to the Doppler tracking measurements, time-domain recordings of the spacecraft's signal at \SI{8.4}{GHz} were also made at PU and GB. These permit measuring the frequency evolution of the spacecraft's reference tone with improved offline processing techniques, such as those developed for spacecraft tracking by the Joint Institute for VLBI ERIC \cite{Calves:2021}, which may allow the observed flicker noise floor to be overcome. Once frequency measurements have been made and their uncertainties estimated, the model and analysis described herein may be applied to determine $\varepsilon$ with improved accuracy. Preliminary work on applying these offline techniques are discussed in \cite{Belonenko:2021} wherein it is estimated that $\sigma_\varepsilon \sim 10^{-5}$ may be attainable. Recordings of RA's signal were also made at other ground radio telescopes for which a partial DCS is possible. Including these in the final solution may allow statistical uncertainties to be further reduced.

For a future space VLBI mission in a highly eccentric orbit, we envision a setup allowing simultaneous recordings of 1-way and 2-way referenced signals in parallel to all downlinks of VLBI recordings. Over a three year period, a mission similar to RA would have $\sim 2500$ sessions, a factor of $20$ increase over the number used in this experiment. Simultaneous recordings would allow a session to be divided into $40$ or more segments, a $10\times$ increase over our average number of segments per session. Further, an orbit with a lower inclination or a tracking station in the southern hemisphere, would allow sessions much closer to perigee increasing the variation in the gravitational redshift by a factor of $10$ or more. Taken together, such a mission could improve the sensitivity of measuring $\varepsilon$ to $\sim 10^{-7}$.

\section{Summary and conclusions}
\label{sec:conclusion}

In this paper we have described a test of the EEP and measurement of the gravitational redshift including:

\begin{enumerate}
    \item details on Doppler-tracking frequency measurements at the PU and GB stations with RA at distances ranging from \SI{25000}{km} to the distance of the Moon,
    \item the implementation of a DCS, similar to GP-A, achieved by alternating RA's communication system between different frequency referencing modes,
    \item the model required to predict relative frequency shifts as small as $10^{-15}$,
    \item measurements of the frequency offset of GB and PU H-masers relative to coordinate time with an accuracy exceeding $10^{-14}$ throughout most of the experiment,
    \item a method for measuring $\varepsilon$ and the SHM frequency offset relative to TCG,
    \item using Monte Carlo simulation to determine the correct weighting of the data using GB and PU noise levels, where in both cases FFM noise was found to dominate,
    \item fit of \SI{8.4}{GHz} data with $\chi^2_\nu = 1.1$ validating the weights derived from simulation,
    \item a final result of $\varepsilon = \left(2.1\pm3.3\right)\times10^{-4}$ using the spacecraft tone at \SI{8.4}{GHz},
    \item the possibility to significantly improve the measurement sensitivity with existing time-domain data, and
    \item the prospect of increasing the sensitivity further with future space VLBI missions to $\sim 10^{-7}$.
\end{enumerate}

\section*{Data availability statement}

The data cannot be made publicly available upon publication due to legal restrictions preventing unrestricted public distribution. The data that support the findings of this study are available upon reasonable request from the authors.

\section*{Acknowledgments}

We thank D.A. Litvinov for his contribution to this experiment. The authors are also grateful to the anonymous reviewers for their useful comments and suggestions. The RadioAstron project is led by the Astro Space Center of the Lebedev Physical Institute of the Russian Academy of Sciences and the Lavochkin Scientific and Production Association under a contract with the Russian Federal Space Agency, in collaboration with partner organizations in Russia and other countries. This paper was supported in part by the Russian Academy of Science Program KP19-270, `The study of the Universe origin and evolution using the methods of earth-based observations and space research.' N.B., M.F.B. and N.V.N. were supported by the National Sciences and Engineering Research Council of Canada.

\section*{References}

\bibliography{cqg23-submit-v4}

\providecommand{\newblock}{}
\begin{thebibliography}{10}
\expandafter\ifx\csname url\endcsname\relax
  \def\url#1{{\tt #1}}\fi
\expandafter\ifx\csname urlprefix\endcsname\relax\def\urlprefix{URL }\fi
\providecommand{\eprint}[2][]{\url{#2}}

\bibitem{Will:2014}
Will C~M 2014 {\em Living Reviews in Relativity\/} {\bf 17} 1--117

\bibitem{Biriukov:2014}
Biriukov A~V, Kauts V~L, Kulagin V~V, Litvinov D~A and Rudenko V~N 2014 {\em
  Astronomy Reports\/} {\bf 58} 783--795

\bibitem{Pound:1960}
Pound R~V and Rebka G~A 1960 {\em Phys. Rev. Lett.\/} {\bf 4}(7) 337--341

\bibitem{Pound:1964}
Pound R~V and Snider J~L 1964 {\em Phys. Rev. Lett.\/} {\bf 13}(18) 539--540

\bibitem{Vessot:1980}
Vessot R~F~C, Levine M~W, Mattison E~M, Blomberg E~L, Hoffman T~E, Nystrom G~U,
  Farrel B~F, Decher R, Eby P~B, Baugher C~R, Watts J~W, Teuber D~L and Wills
  F~D 1980 {\em Phys. Rev. Let.\/} {\bf 45} 2081--2084

\bibitem{Vessot:1989}
{Vessot} R~F~C 1989 {\em Advances in Space Research\/} {\bf 9} 21--28

\bibitem{Delva:2018}
Delva P, Puchades N, Sch{\"o}nemann E, Dilssner F, Courde C, Bertone S,
  Gonzalez F, Hees A, Le~Poncin-Lafitte C, Meynadier F {\em et~al.\/} 2018 {\em
  Phys. Rev. Let.\/} {\bf 121} 231101

\bibitem{Herrmann:2018}
Herrmann S, Finke F, L{\"u}lf M, Kichakova O, Puetzfeld D, Knickmann D, List M,
  Rievers B, Giorgi G, G{\"u}nther C {\em et~al.\/} 2018 {\em Phys. Rev.
  Let.\/} {\bf 121} 231102

\bibitem{Takamoto:2020}
{Takamoto} M, {Ushijima} I, {Ohmae} N, {Yahagi} T, {Kokado} K, {Shinkai} H and
  {Katori} H 2020 {\em Nature Photonics\/} {\bf 14} 411--415

\bibitem{Savalle:2019}
Savalle E, Guerlin C, Delva P, Meynadier F, le~Poncin-Lafitte C and Wolf P 2019
  {\em Classical and Quantum Gravity\/} {\bf 36} 245004

\bibitem{Litvinov:2021b}
Litvinov D and Pilipenko S 2021 {\em Classical and Quantum Gravity\/} {\bf 38}
  135010

\bibitem{Kardashev:2013}
Kardashev N~S, Khartov V~V, Abramov V~V, Avdeev V~Y, Alakoz A~V, Aleksandrov
  Y~A, Ananthakrishnan S, Andreyanov V~V, Andrianov A~S, Antonov N~M {\em
  et~al.\/} 2013 {\em Astronomy Reports\/} {\bf 57} 153--194

\bibitem{Alexandrov:2012}
{Alexandrov} Y~A, {Andreyanov} V~V, {Babakin} N~G, {Babyshkin} V~E, {Belousov}
  K~G, {Belyaev} A~A, {Biryukov} A~V, {Bubnov} A~E, {Bykadorov} A~A,
  {Vasil'kov} V~I, {Vinogradov} I~S, {Gvamichava} A~S, {Zinoviev} A~N, {Komaev}
  R~V, {Kanevskiy} B~Z, {Kardashev} N~S, {Kovalev} Y~A, {Kovalev} Y~Y,
  {Kovalenko} A~V, {Korneev} Y~A, {Kostenko} V~I, {Kreisman} B~B, {Kukushkin}
  A~Y, {Larionov} M~G, {Likhachev} S~F, {Likhacheva} L~N, {Medvedev} S~Y,
  {Melekhin} M~V, {Mizyakina} T~A, {Nikolaev} N~Y, {Novikov} B~S, {Novikov}
  I~D, {Pavlenko} Y~K, {Ponomarev} Y~N, {Popov} M~V, {Pyshnov} V~N, {Rozhkov}
  V~M, {Sakharov} B~A, {Serebrennikov} V~A, {Smirnov} A~I, {Stepanyants} V~A,
  {Fedorchuk} S~D, {Shatskaya} M~V, {Sheikhet} A~I, {Shirshakov} A~E and
  {Yakimov} V~E 2012 {\em Solar System Research\/} {\bf 46} 458--465

\bibitem{Nunes:2020}
Nunes N~V, Bartel N, Bietenholz M~F, Zakhvatkin M~V, Litvinov D~A, Rudenko V~N,
  Gurvits L~I, Granato G and Dirkx D 2020 {\em Advances in Space Research\/}
  {\bf 65} 790 -- 797

\bibitem{Litvinov:2018}
Litvinov D~A, Rudenko V~N, Alakoz A~V, Bach U, Bartel N, Belonenko A~V,
  Belousov K~G, Bietenholz M, Biriukov A~V, Carman R, Cim{\'o} G, Courde C,
  Dirkx D, Duev D~A, Filetkin A~I, Granato G, Gurvits L~I, Gusev A~V, Haas R,
  Herold G, Kahlon A, Kanevsky B~Z, Kauts V~L, Kopelyansky G~D, Kovalenko A~V,
  Kronschnabl G, Kulagin V~V, Kutkin A~M, Lindqvist M, Lovell J~E~J, Mariey H,
  McCallum J, Molera~Calv{\'e}s G, Moore C, Moore K, Neidhardt A, Pl{\"o}tz C,
  Pogrebenko S~V, Pollard A, Porayko N~K, Quick J, Smirnov A~I, Sokolovsky K~V,
  Stepanyants V~A, Torre J~M, de~Vicente P, Yang J and Zakhvatkin M~V 2018 {\em
  Physics Letters A\/} {\bf 382} 2192--2198

\bibitem{Zakhvatkin:2020}
Zakhvatkin M~V, Andrianov A~S, Avdeev V~Y, Kostenko V~I, Kovalev Y~Y, Likhachev
  S~F, Litovchenko I~D, Litvinov D~A, Rudnitskiy A~G, Shchurov M~A {\em
  et~al.\/} 2020 {\em Advances in Space Research\/} {\bf 65} 798--812

\bibitem{Crouse:2017}
Crouse D~F 2017 {\em IEEE Aerospace and Electronic Systems Magazine\/} {\bf 32}
  18--27

\bibitem{Pavlis:2012}
Pavlis N~K, Holmes S~A, Kenyon S~C and Factor J~K 2012 {\em Journal of
  Geophysical Research: Solid Earth\/} {\bf 117} B04406

\bibitem{IERS2010}
Petit G and Luzum B 2010 {\em IERS Conventions 2010\/} vol~36

\bibitem{Folkner:2009}
Folkner W~M, Williams J~G and Boggs D~H 2009 {\em IPN Progress Report\/} {\bf
  42} 1--34

\bibitem{Landskron:2018vmf3}
Landskron D and B{\"o}hm J 2018 {\em Journal of Geodesy\/} {\bf 92} 349--360

\bibitem{Landskron:2018grad}
Landskron D and B{\"o}hm J 2018 {\em Journal of Geodesy\/} {\bf 92} 1387--1399

\bibitem{Noll:2010}
Noll C~E 2010 {\em Advances in Space Research\/} {\bf 45} 1421--1440

\bibitem{Litvinov:2021}
Litvinov D~A, Nunes N~V, Filetkin A~I, Bartel N, Gurvits L~I, Molera~Calves G,
  Rudenko V~N and Zakhvatkin M~V 2021 {\em Advances in Space Research\/} {\bf
  68} 4274--4291

\bibitem{Vessot:2005}
Vessot R~F~C 2005 {\em Metrologia\/} {\bf 42} S80

\bibitem{Barnes:1971}
Barnes J~A, Chi A~R, Cutler L~S, Healey D~J, Leeson D~B, McGunigal T~E, Mullen
  J~A, Smith W~L, Sydnor R~L, Vessot R~F~C and Winkler G~M~R 1971 {\em IEEE
  Transactions on Instrumentation and Measurement\/} {\bf IM-20} 105--120

\bibitem{Zucca:2005}
Zucca C and Tavella P 2005 {\em IEEE Transactions on Ultrasonics,
  Ferroelectrics, and Frequency Control\/} {\bf 52} 289--296

\bibitem{Xu:2019}
Xu C 2019 {\em The Astronomical Journal\/} {\bf 157} 127

\bibitem{Damour:2012}
Damour T 2012 {\em Classical and Quantum Gravity\/} {\bf 29} 184001

\bibitem{Calves:2021}
{Molera Calv{\'e}s} G, {Pogrebenko} S~V, {Wagner} J~F, {Cim{\`o}} G, {Gurvits}
  L~I, {Bocanegra-Baham{\'o}n} T~M, {Duev} D~A and {Nunes} N~V 2021 {\em
  Publications of the Astronomical Society of Australia\/} {\bf 38} e065

\bibitem{Belonenko:2021}
{Belonenko} A~V, {Gusev} A~V and {Rudenko} V~N 2021 {\em Gravitation and
  Cosmology\/} {\bf 27} 383--391

\end{thebibliography}

\end{document}